# Magnetic anisotropy of large floating-zone-grown single-crystals of SrRuO$_3$

S. Kunkemöller1, F. Sauer1, A. A. Nugroho2, and M. Braden*,1

SrRuO$_3$ is a highly interesting material due to its anomalous-metal properties related with ferromagnetism and its relevance as conductive perovskite layer or substrate in heterostructure devices. We have used optical floating zone technique in an infrared image furnace to grow large single crystals of SrRuO$_3$ with volumes attaining several hundred mm$^3$. Crystals obtained for optimized growth parameters exhibit a high ferromagnetic Curie temperature of 165 K and a low-temperature magnetization of 1.6 $\mu_B$ at a magnetic field of 6 T. The high quality of the crystals is further documented by large residual resistance ratios of 75 and by crystal structure and chemical analyzes. With these crystals the magnetic anisotropy could be determined.

## 1 Introduction

SrRuO$_3$ is the infinite-layer material of the Ruddlesden-Popper series of ruthenates Sr$_{n+1}$Ru$_n$O$_{3n}$, which have attracted enormous interest mostly due to the unconventional superconductivity in Sr$_2$RuO$_4$, which is thought to be related to a ferromagnetic instability [1, 2]. In addition to unconventional superconductivity, perovskite ruthenates exhibit metamagnetic transitions, Mott insulating phases, spin-density wave ordering arising from nesting and hidden order phases, to cite the most prominent features [3, 4, 5, 6].

SrRuO$_3$ is particularly interesting as it is the only simple material (otherwise one has to study three or higher-number layer members of the Ruddlesden-Popper series) that exhibits ferromagnetic order at ambient conditions [7], which still is a rare phenomenon in the broader class of transition metal oxides. Its ferromagnetism inspired the proposal of p-wave superconductivity mediated through magnetic fluctuations in Sr$_2$RuO$_4$ [1, 2] and it is coupled to the quantum-phenomena associated with the metamagnetic transitions in double and single layered materials [3, 4]. The magnetic moment of ferromagnetic SrRuO$_3$ amounts to 1.6 $\mu_B$ and the ferromagnetic order occurs near 160 K [7]. Ferromagnetism in SrRuO$_3$ is associated with strongly anomalous behavior in various properties: Thermal expansion shows an invar effect in the ferromagnetic phase suggesting a change in the local versus itinerant character of magnetic moments [8, 9]. Furthermore, ferromagnetic correlations clearly interfere with the electronic transport yielding a linear resistivity that breaks the Ioffe rule already at 500 K [10]. Further interest in this material arises from its suitability as a conducting and magnetic layer in various perovskite oxide heterostructures [9]. With the possibility to grow large crystals, SrRuO$_3$ could even be used as a substrate.

In 1959 Randall and Ward reported the first synthesis of polycrystalline SrRuO$_3$ [11]. Single crystals grown by a flux method yield masses in the mg regime [12], but the purity of these SrRuO$_3$ crystals is not comparable with that of other layered ruthenates crystals grown with the floating zone technique [13]. Ikeda et al. reported attempts to grow also SrRuO$_3$ with the floating zone method, but no single crystals could be obtained due to difficulties in stabilizing the floating zone [14]. Very recently Kikugawa et al. re-ported about the successful growth of single crystals in the floating zone [15].

Table 1 Summary of various growth parameters and their effect on the single-crystal growth of SrRuO$_3$. The optimized parameters for the crystal growth of SrRuO$_3$ are given in the last row.

| parameter | variation | remarks | optimum |
|---|---|---|---|
| Sintering | O$_2$, 1000°C, 12 hC | High evaporation of RuO$_2$ | Air, 1350°C, 3 h |
| **Feed rod** RuO$_2$ excess Ø, RuO$_2$ excess | 60 % 1 cm, 60 % | 327 and 214 phase Crystals with minor quality | 90 % 0.7 cm, 90 % |
| Gas flow | 1.5 l/min | More RuO$_2$ deposition on the glass tube | 4.5 l/min |
| Atmosphere | 100 % O$_2$ 40 % O$_2$ 25 % O$_2$ | No stable molten zone Unstable molten zone High melting ruthenium metal phase | 33 % O$_2$ |
| Growth rate | 17 mm/h 10 mm/h 7 mm/h | Small crystallites 327 and 214 phase 214 and 327 phase | 15 mm/h |

Here we also present a detailed description of the successful single-crystal growth of SrRuO$_3$ using the optical floating zone technique [16]. The parameters we used for the crystal growth strongly differ from the ones used by Kikugawa et al. [15]. Most importantly, we used more than the double speed for the crystal growth and oxygen content. Therefor crystals with a mass of up to 3 g could be obtained for optimized growth parameters. These crystals were characterized by several microscopic and macroscopic methods. In addition we present a full structural characterization by several x-ray scattering techniques, which allows us to determine the complex twinning of that perovskite. We also describe detwinning of SrRuO$_3$ crystals, which gives access to the anisotropic magnetic properties.

## 2 Single-Crystal growth in SrRuO$_3$ in a mirror furnace

### 2.1 General procedure of the crystal growth of SrRuO$_3$

The parameters to obtain large high quality crystals had to be determined by stepwise variation following a common general procedure. The detailed parameters and their effect on the growth are summarized in table 1. SrCO$_3$ and RuO$_2$ powders both with minimum purity of 99.95 % were mixed with a molar ratio of 1:1.6 or 1:1.9 and homogenized by ball-mixing. A preliminary reaction was performed by heating the powder in a platinum crucible at 1000 °C for 24 hours in air with an intermediate grinding. The feed rod was fabricated by pressing the powder in a hydraulic press with a pressure of 1500 kg/cm$^2$. The resulting rod was sintered for several hours at 1000 °C or 1350 °C trying different atmospheres. Powder x-ray diffraction (XRD) measurements show that the polycrystalline materials obtained in the preliminary reaction and in the sintered rod consist of a mixture of SrRuO$_3$ and excess RuO$_2$.

For the single-crystal growth an infrared image furnace Canon Machinery Inc. SC1-MDH11020-CE equipped with two 2000 W halogen lamps and a cold trap was used. The atmosphere was varied by mixing Argon and Oxygen while steadily applying the maximum available total pressure of 10 bar and a gas flow of 1.5 l/min or 4.5 l/min. The crystals were grown with a speed varying between 7 mm/h and 17 mm/h, the speed of the feed rod was set to twice that speed. Both shafts rotated with 15 rpm in opposite directions.

Crystals grown with an optical floating zone technique are usually limited in length by the restrictions of the furnace used, such as the possibility to move the feed and seed rod with respect to the lamps. In contrast, the crystal size of SrRuO$_3$ is essentially limited by the volatility of RuO$_2$. The evaporated material condenses at the glass tube absorbing the incoming light. To compensate for this absorption the power of the lamps has to be continuously enhanced until reaching the maximum power or substantial heating of the glass tube, which can cause serious damage to the furnace. A cold trap is used to absorb the evapo-rated material, but the trap becomes rapidly saturated. After evaporation of about 2 g of $RuO_2$, we observe that the material is no longer fixed at the cold trap but rapidly covers the glass tube and blocks the heating. Therefore, the growth parameter had also to be optimized with respect to the amount of evaporated $RuO_2$ per volume of single-crystalline material.

## 2.2 Optimization of growth parameters

For the preparation of the feed rod we tried diameters of 0.7 cm and 1 cm with excess of $RuO_2$ of 60 % and 90 %. With the thicker feed rod an excess of 60 % turned out to be sufficient to get single crystals as the larger diameter slightly reduces the evaporation problem. But the quality of the crystals obtained with this larger diameter was significantly below that of those obtained with the thinner diameter. The minor crystalline quality is observed as smeared spots of the Laue images, and the saturation magnetic moment of only 1.55 $\mu_B$ and the RRR of 16 are well beyond the values found for the optimized crystals.

The pressure was always set to 10 bar, the highest one applicable to the furnace, since for high pressure less evaporation is expected. Also the rotation speeds of both shafts of 15 rpm were not varied, because they are expected to have little effect on the crystal growth.

The optimum growth speed was found to be 15 mm/h. For the higher speed of 17 mm/h only small single crystalline grains could be obtained even though the molten zone was as stable as in the attempts with optimum speed. In the attempts with lower growth speed, 10 mm/h and 7 mm/h, respectively, the resulting crystals consist of a mixture of $Sr_3Ru_2O_7$ (327) and $Sr_2RuO_4$ (214) with more $Sr_2RuO_4$ in the attempt with the lower speed. This shows that more $RuO_2$ evaporates during the growth processes with slower growth speed, which needs to be compensated by a greater excess of $RuO_2$ in the feed rod.

The constitution of the atmosphere has strong impact on the stability of the molten zone. For a high oxygen content of 40 % and 100 % it was not possible to keep the molten zone stable for a sufficiently long period to grow single crystalline material. In several attempts with lower oxygen partial pressure (only 25 %) a Ru metal phase formed, and the crystal growth was not stable. The formation of this metallic phase could be suppressed by stopping the motion of the feed rod. This indicates that this phase occurs only in Ru rich

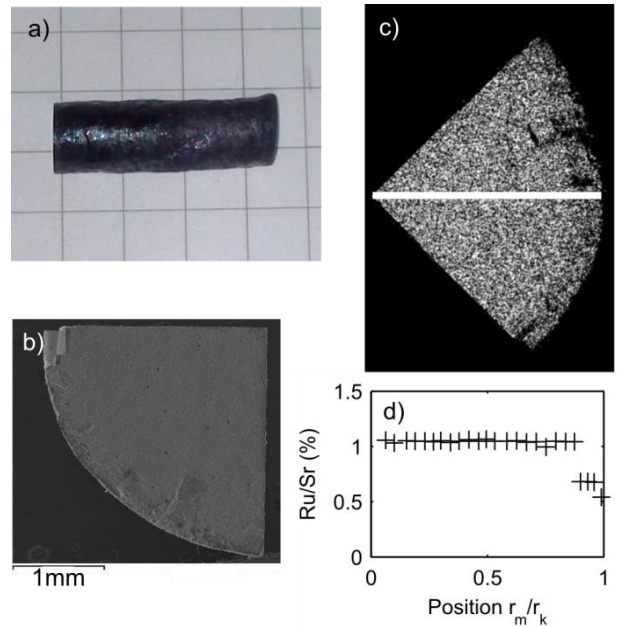

**Figure 1** (a) Single crystalline part of a growth attempt. (b) Electron microscopy picture of a quarter of a slice cut perpendicular to the rod. (c) Mapping of the Ru content with EDX analysis of the piece shown in (b). (d) Ratio of Sr to Ru contents along the white marked path in (c).

floating zones. The $O_2$ reduced atmosphere suppresses thus the evaporation of $RuO_2$, however, it is much more difficult to keep the floating zone stable under these conditions. The higher oxygen partial pressure clearly results in larger crystals and makes the crystal growth more reproducible.

With a total gas flow of 1.5 l/min the power of the lamps had to be increased by 6 % during the crystal growth to compensate for the absorption of light by the material deposited on the glass tube. With a higher gas flow of 4.5 l/min electric power had to be increased by only 1 % for the same growth length in order to keep the molten zone stable. Applying a higher gas flow is technically not possible.

## 3 Chemical and structural analysis of the optimized single crystals

In Figure 1 the single crystalline part of a successful crystal growth can be seen in part a). As-grown crystals have a thick polycrystalline skin. An electron microscopy picture of a quarter of a slice can be seen in b). This piece was further investigated by electron dispersive x-ray spectroscopy (EDX). The obtained mapping of the Ru content can be seen in panel c). This picture reveals an inner bulk part with the cor-

**Table 1** Crystal structure of SrRuO$_3$ at room temperature. The atomic positions are given in fractions of the unit cell, the atomic displacements are given in Å$^2$, values in brackets indicates the error on the last digits

|  | Sr | Ru | O1 | O2 |
|---|---|---|---|---|
| X | 0.01676(5) | 0 | 0.4972(3) | 0.2769(3) |
| Y | 0.25 | 0 | 0.25 | 0.0273(2) |
| Z | -0.00256(7) | 0,5 | 0.0544(4) | 0.7235(3) |
| $U_{11}$ | 0.00663(14) | 0.00147(10) | 0.0153(14) | 0.0115(8) |
| $U_{22}$ | 0.00455(16) | 0.00168(11) | 0.0011(11) | 0.0077(8) |
| $U_{33}$ | 0.00697(16) | 0.00355(11) | 0.0067(11) | -0.0019(8) |
| $U_{12}$ | 0 | -0.00009(10) | 0 | -0.0019(6) |
| $U_{13}$ | -0.00102(13) | 0.00030(13) | 0.0009(7) | -0.0045(6) |
| $U_{23}$ | 0 | 0.00009(11) | 0 | 0.0011(6) |

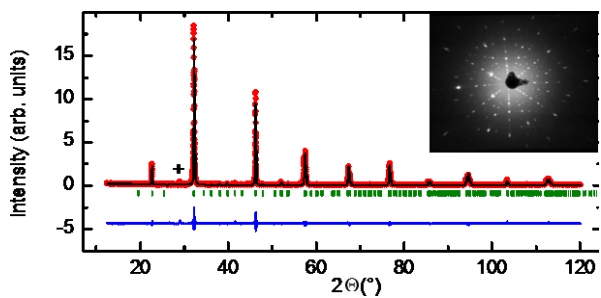

**Figure 2** XRD powder pattern of a crushed single crystal. The red circles denote the data points, the black line is the Rietveld fit, the green marks indicate the peak positions according to space group Pnma and the blue line is the difference between the data and the fit. The asterisk marks the position of the strongest RuO$_2$ peak. The inset shows a Laue image of a single crystal.

rect Ru content and a skin containing less Ru. The white line indicates the positions of EDX measurements, whose results are shown in d) as a function of the distance from the center of the rod, $r_m$, scaled to the total radius of the crystal, $r_k$. These EDX results confirm the existence of a bulk SrRuO$_3$ crystal containing the correct amount of Ru covered by a skin with a total volume of about one quarter of the entire cylinder. The skin can be easily removed by filing. The EDX analysis of the bulk yields an average ratio of ruthenium to strontium of 1.044(3). The powder XRD measurements were performed with Cu K$\alpha$ radiation on crushed single crystals. Rietveld fits were performed with the FulProf Suite [17] in the orthorhombic space group Pnma with the structural data taken from our single crystal XRD measurements, see below. The powder diffraction studies show that the bulk crystals are indeed SrRuO$_3$ with a very small inclusion (~3 %) of a RuO$_2$ impurity phase (see Figure 2). The position of the strongest RuO$_2$ peak is marked in the pattern. The expected ratio of ruthenium to strontium for crystals including this impurity phase is greater than one which is supported by the EDX analysis. Powder patterns of the skin show a mixture of the layered Strontium Ruthenates, especially the single and double-layered materials. The lattice constants determined by our powder XRD studies are a=5.5322(1) Å, b=7.8495(2) Å and c=5.5730(1) Å in good agreement with previously published neutron powder measurements [18, 19]. Laue images (see inset Figure 2) taken at different positions at the surface of the crystals indicate single crystallinity. For a complete crystal structure determination we used a single crystal X-ray diffractometer X8-APEX by Bruker AXS with a goniometer in kappa-geometry and x-ray radiation from a molybdenum anode. The wavelength was $\lambda=0.71073$ Å and the distance between the sample and the detector was set to 50 mm. The analysis of several crystals shows that even small

SrRuO$_3$ crystals with dimensions below 0.1 mm are twinned with non-equal twinning fractions. In the structure refinement the 6 possible orientations of twin domains were taken into account. Structure refinements were carried out using Jana2006 [20]. 20720 observations were merged into 3801 Bragg intensities by averaging only identical and Friedelequivalent reflections. The data were corrected for absorption, and a type I extinction correction was applied during the refinements. A good description of the experimental intensities was achieved yielding weighted reliability-values of $Rw$=3.08 % and 3.82 % for the observed (larger than three $\sigma$- values) and all reflections, respectively. The goodness of fit value amounts to 1.27 for the observed reflections. The positional and anisotropic displacement parameters from our structure refinements are given in Table 2. Most of the positional parameters agree within the error bars with a previous powder neutron diffraction study [19], just the minor difference of the $z$ parameter of the O1 site seems to arise from the refinement of isotropic displacement parameters in the powder study. Note, that the precision of the parameters is significantly higher in our single-crystal analysis, even for the light O sites. The refinement of the occupation of the strontium to ruthenium ratio yields 1.0009(12) indicating perfect stoichiometry of the crystals obtained with the optimized growth conditions.

The crystal structure of SrRuO$_3$ differs from the ideal cubic perovskite structure by a rotation of the RuO$_6$-octahedra around the $b$-axis (long axis in space group Pnma), combined with a tilt [18, 22]. Here, the corresponding angles are called "rotation angle" and "tilt angle", respectively. The rotation angle and the tilt angle are not defined unambiguously since the octahedra are slightly distorted. In order to determine

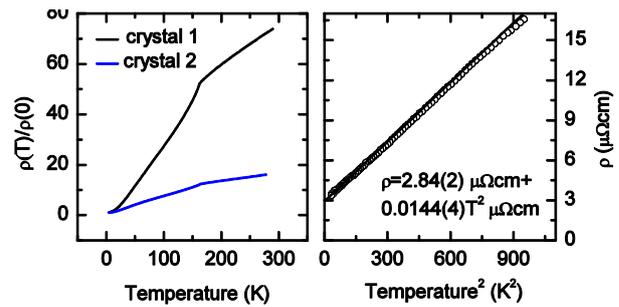

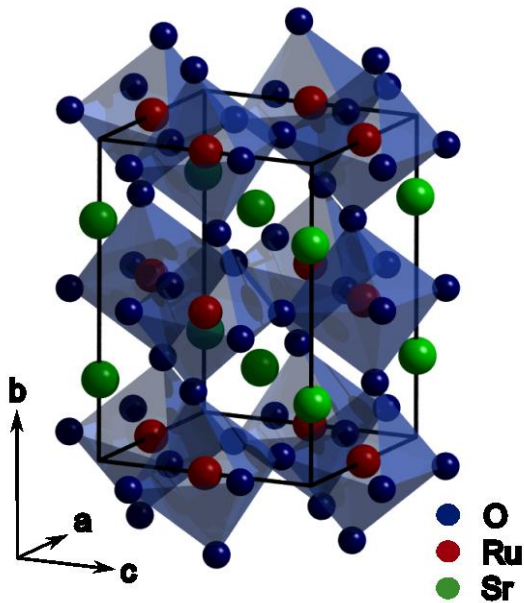

**Figure 2** Drawing of the crystal structure of $SrRuO_3$ as obtained by the single-crystal X-ray diffraction experiment at room temperature, see table 1

the rotation angle $\varphi$, the atom O2 is projected onto the $ac$-plane and a Ru – Ru – O2′ angle is measured yielding $\varphi=6.10(3)$ °. The tilting angle can be measured via the displacement of the apical oxygen O1, $\Theta_{ap}=8.79(5)$ ° and via the tilt of the basal plane, $\Theta_{bas}=8.72(5)$ °. The space group $Pnma$ also allows for a deformation of the $RuO_6$ octahedron which can be associated with orbital polarization of ferro- or antiferro-orbital type [22]. In $SrRuO_3$ there is no significant splitting of the two Ru - O2 distances of 1.9858 Å and 1.9859 Å, also the Ru-O1 distance is almost identic, but the basal plane of the octahedron is distorted, the distance of the octahedron edge length along $a$ O2 - O2$_a$=2.782 Å is significantly smaller than that along $c$ O2 - O2$_c$=2.835 Å. A qualitatively similar but much larger splitting is observed in insulating $Ca_2RuO_4$ [23]. This deformation is related with the fact that the lattice is shorter along the tilt axis, which contrasts with a rigid picture, but which is observed in many transition-metal oxides [22].

## 4. Characterization of physical properties

Resistivity measurements on our crystals are presented in Figure 4. In a) the temperature dependent resistivity divided by the low-temperature residual resistivity measured with a standard four-point method between 5 K and room temperature is shown.

**Figure 3** (a) Temperature dependent resistivity divided by the residual resistivity. The black and blue lines correspond to the measurements on a crystal with optimized and non-optimized growth conditions, respectively. (b) Resistivity plotted against the squared temperature for the crystal with optimized growth parameters. The black line is a linear fit of the data between 5 K and 10 K

The results are in good agreement with the data presented in reference [21]; in particular we find the clear kink in the resistivity at the ferromagnetic phase transition. The residual resistivity ratio obtained for the crystal grown under optimized conditions amounts to 75, and is thus larger than that in the cleanest $SrRuO_3$ films [24]. Furthermore, the residual resistivity of $\rho_0=3$ $\mu\Omega$cm is less than half of the value observed in flux grown single crystals [12]. In contrast a crystal grown with non-optimum conditions (feed rod diameter 1 cm, $RuO_2$ excess 60 %) exhibits a much lower residual resistivity ratio, although a high ferromagnetic $T_c$ is clearly visible. In b) the specific resistivity is plotted against the squared temperature, and the data between 5 K and 10 K are fitted with a linear function indicating Fermi liquid behavior. From this fit the residual resistivity and the $A$ coefficient were determined: $\rho=\rho_0+A\cdot T^2$ with $A=0.0144(4)$ $\mu\Omega$ cm/K$^2$ and $\rho_0=2.84(2)$ $\mu\Omega$ cm. Note that the errors do not include uncertainties in the geometry. Magnetization measurements (Figure 5) were performed with a commercial SQUID Magnetometer (Quantum Design) and show the same characteristics of the ferromagnetic order as the data from reference [7]. The magnetization loop in a) is measured at 5 K with the field applied along a pseudocubic [110] direction (note that our crystals are twinned). The magnetic moment does not saturate in the applied field of 6 T and reaches values of 1.6 $\mu_B$. The coercive field amounts only to 14(2) Oe. The ferromagnetic phase transition occurs at 165 K as it can be seen in panel b), which shows the temperature dependent magnetization measured in a magnetic field of 50 Oe applied along a pseudocubic [110] direction measured on cooling. Kikugawa et al. report a lower magnetic quality, a maximum magnetic moment of about 1.4 $\mu_B$ in a magnetic field of 7 T and a coercive field of 40 Oe [15] most likely because of mixing of various directions.

* M. Braden: e-mail: braden@ph2.uni-koeln.de, phone: +49 221 470 3655

[1] II. Physikalisches Institut, Universität zu Köln, Zülpicher Straße 77, D-50937 Köln, Germany

[2] *Faculty of Mathematics and Natural Science, Institut Teknologi Bandung, Jalan Ganesha 10, 40132 Bandung, Indonesia*

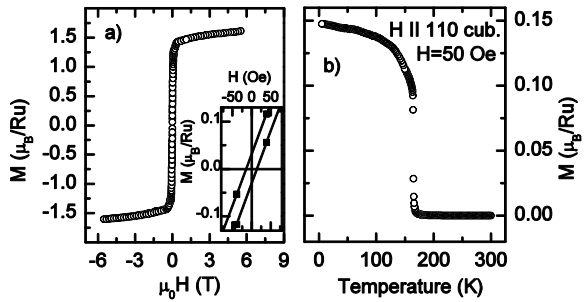

**Figure 4** Magnetic characterization. The magnetic field is applied along a pseudo cubic [110] direction. (a) Magnetization curve measured at 5 K. The inset shows the coercive field of only 14 Oe. (b) Temperature dependent magnetization measured upon cooling in 50 Oe. The magnetic phase transition occurs at 165 K.

## 5. Anisotropic magnetic properties

In order to detwin the crystals uniaxial pressure of 1.5 kg/mm$^2$ were applied on a rectangular crystal on a cubic (110) face heated to 730 °C and furnace cooled to room temperature. With the single-crystal x-ray diffractometer several unique reflexions (e.g. (850)) for the twins were measured in reflection geometry confirming the successful and complete detwinning of the crystal.

Magnetic hysteresis loops recorded at 5K are shown in Fig. 5 for magnetic fields applied along the three orthorhombic axes, which are accessible with the untwinned crystal. There is little anisotropy between the orthorhombic a and c axes although these lattice constants and the O2-O2 octahedron edges differ considerably. In contrast, the magnetization along the **b** direction is strongly reduced. The fact that anisotropy persists to large fields is remarkable and points to strong spin-orbit coupling in SrRuO$_3$. Along the **b** direction magnetization rapidly increases to ~0.5 $\mu_B$ but then grows with a small field slope. This behavior indicates that the directions parallel to the O-O edges of the RuO6 octahedron are softer than those along the bonds. A first principles study should be able to shed further light on this issue. The magnetic anisotropy reported by Kikugawa et al. [15] is much smaller than what we observe, most likely due to twinning, and an early study [25] finds qualitatively a similar anisotropy but with ~30\% smaller saturation magnetization most likely due to lower quality of flux-grown samples.

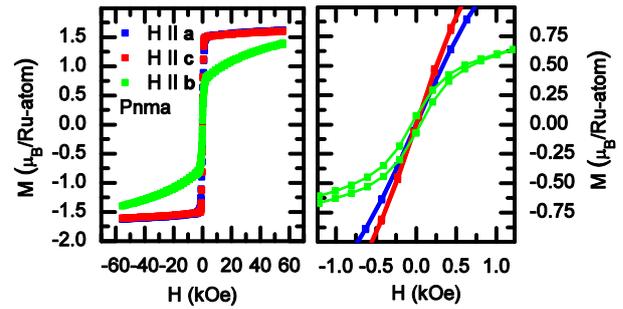

**Figure 5** Magnetic anisotropy. (a) Magnetization curves measured at 5 K. The magnetic field is applied along the three principal symmetry directions. (b) A zoom into panel (a).

## 6. Conclusion

We present the crystal growth of large SrRuO$_3$ single crystals with the optical floating zone technique. Crystals with a mass of up to 3 g (several hundred mm$^3$ volume) could be obtained. The residual resistivity ratio and the small coercive field as well as chemical analyzes denote the high purity of the crystals. The single-crystal XRD analysis confirms the ideal stoichiometry of these crystals and reveals that even sub-mm sized crystals exhibit complex twinning with six possible domain orientations. The magnetic anisotropy of untwinned crystals is shown.

**Acknowledgements**.

Part of this work was funded through the Institutional Strategy of the University of Cologne within the German Excellence Initiative. We thank T. Fröhlich for fruitful discussions.

## References

[1] Y. Maeno, H. Hashimoto, K. Yoshida, S. Nishizaki, T. Fujita, J. G. Bednorz, F. Lichtenberg, Nature (London) **372,** 532–534 (1994).
[2] T. M. Rice, M. Sigrist, J. Phys.: Cond. Matter **7,** L643–L648 (1995)
[3] R. S. Perry, L. M. Galvin, S. A. Grigera, L. Capogna, A. J. Schofield, A. P. Mackenzie, M. Chiao, S. R. Jul-

ian, S. I. Ikeda, S. Nakatsuji, Y. Maeno, C. Pfleiderer, Phys. Rev. Lett. **86,** 2661–2664 (2001).
[4] S. Nakatsuji, D. Hall, L. Balicas, Z. Fisk, K. Sugahara, M. Yoshioka, Y. Maeno, Phys. Rev. Lett. **90,** 137202 (2003).
[5] S. Nakatsuji, T. Ando, Z. Mao, Y. Maeno, Physica B **259-261,** 949–950 (1999).
[6] M. Braden, O. Friedt, Y. Sidis, P. Bourges, M. Minakata, Y. Maeno, Phys. Rev. Lett. **88,** 197002 (2002).
[7] A. Kanbayasi, Journal of the J. Phys. Soc. Jpn. **41,** 1876–1878 (1976).
[8] T. Kiyama, K. Yoshimura, K. Kosuge, Y. Ikeda, Y. Bando Phys. Rev. B **54,** R756–R759 (1996).
[9] G. Koster, L. Klein, W. Siemons, G. Rijnders, J. S. Dodge, C.-B. Eom, D. H. A. Blank, M. R. Beasley, Rev. Mod. Phys. **84 ,** 253–298 (2012).
[10] L. Klein, J. S. Dodge, C. H. Ahn, G. J. Snyder, T. H. Geballe, M. R. Beasley, A. Kapitulnik, Phys. Rev. Lett. **77,** 2774–2777 (1996).
[11] J. J. Randall, R. Ward, J. Am. Chem. Soc. **81,** 2629–2631 (1959).
[12] R. Bouchard, J. Gillson, Mater. Res. Bull. **7,** 873–878 (1972).
[13] L. Capogna, A. P. Mackenzie, R. S. Perry, S. A. Grigera, L. M. Galvin, P. Raychaudhuri, A. J. Schofield, C. S. Alexander, G. Cao, S. R. Julian, Y. Maeno, Phys. Rev. Lett. **88,** 076602 (2002).
[14] S. I. Ikeda, U. Azuma, N. Shirakawa, Y. Nishihara, Y. Maeno, J. Cryst. Growth **237-239,** 787–791 (2002).
[15] N. Kikugawa, R. Baumbach, J. S. Brooks, T. Terashima, S. Uji, Y. Maeno, Cryst. Growth Des. **15,** 5573-5577 (2015)
[16] H. A. Dabkowska and A. B. Dabkowski, Springer Handbook of Crystal Growth, G. Dhanaraj, K. Byrappa, V. Prasad, M. Dudley, Springer Verlag, Berlin Heidelberg (2010).
[17] J. Rodríguez-Carvajal, Physica B **192,** 55–69 (1993).
[18] C. W. Jones, P. D. Battle, P. Lightfoot, W. T. A. Harrison, Acta Crystallogr. **45,** 365–367 (1989).
[19] S. N. Bushmeleva, V. Yu. Pomjakushin, E. V. Pomjakushina, D. V. Sheptyakov, A. M. Balagurov, J Magn. Magn. Mater. **305,** 491–496 (2006).
[20] V. Petříček, M. Dušek, L. Palatinus, Z. Kristallogr. **229,** 345–352 (2004)
[21] P. B. Allen, H. Berger, O. Chauvet, L. Forro, T. Jarlborg, A. Junod, B. Revaz, G. Santi, Phys. Rev. B **53,** 4393–4398 (1996).
[22] M. Cwik, T. Lorenz, J. Baier, R. Müller, G. André, F. Bourée, F. Lichtenberg, A. Freimuth, R. Schmitz, E. Müller-Hartmann, M. Braden, Phys. Rev. B **68,** 060401(R) (2003).
[23] M. Braden, G. André, S. Nakatsuji, Y. Maeno Phys. Rev. B **58,** 847 (1998).
[24] A. P. Mackenzie, J. W. Reiner, A. W. Tyler, L. M. Galvin, S. R. Julian, M. R. Beasley, T. H. Geballe, A. Kapitulnik, Phys. Rev. B **58,** R13318–R13321 (1998).
[25] G. Cao, S. McCall, M. Shepard, J. E. Crow, Phys. Rev B **56,** 321-329 (1997)

Table of Contents

Page xx-yy (for typesetter)

S. Kunkemöller, F. Sauer, A. A. Nugroho, and M. Braden

**Magnetic anisotropy of large floating-zone-grown single-crystals of SrRuO$_3$**

The magnetocrystalline anisotropy of SrRuO$_3$ of large floating zone crystals is presented. The anisotropy is only accessible in untwinned crystals. We describe the crystal growth process and the untwinning procedure. Our single crystal x-ray diffraction analysis allows us to investigate the complex twinning and to determine the different orthorhombic directions of our SrRuO$_3$ single-crystals.

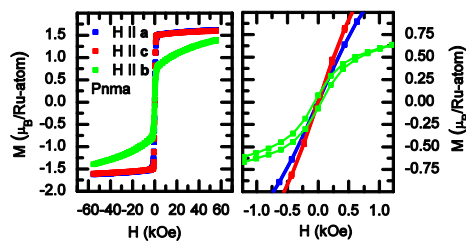

Original Papers